\documentclass[a4paper,11pt]{article}
\pdfoutput=1
\usepackage{jheppub}
\usepackage{graphicx,slashed,hyperref,color,multirow}
\usepackage{amssymb}
\usepackage[normalem]{ulem}
\usepackage[utf8]{inputenc}
\usepackage{braket}
\usepackage{pifont}
\hypersetup{
   bookmarks=true,         
   unicode=true,          
   pdftoolbar=true,        
   pdfmenubar=true,        
   pdffitwindow=false,     
   pdfstartview={FitH},    
   pdftitle={My title},    
   pdfauthor={Author},     
   pdfsubject={Subject},   
   pdfcreator={Creator},   
   pdfproducer={Producer}, 
   pdfkeywords={keyword1} {key2} {key3}, 
   pdfnewwindow=true,      
   colorlinks=true,       
   linkcolor=blue,          
   citecolor=magenta,        
   filecolor=magenta,      
   urlcolor=cyan           
}

\usepackage{amssymb}
\usepackage{enumerate}
\usepackage{amssymb}
\usepackage{amsmath}
\usepackage{tikz}
\usepackage[compat=1.1.0]{tikz-feynman}
\usepackage{feynmf}
\usepackage{makecell}
\usepackage{slashed}
\usepackage{natbib}
\usepackage{graphicx}
\hypersetup{colorlinks=true}

\usepackage{mathrsfs,amssymb}  
\usepackage{cancel}
\usepackage[normalem]{ulem}

\newcommand\be{\begin{equation}}
\newcommand\ee{\end{equation}}
\newcommand{\comment}[1]{}
\newcommand\bea{\begin{eqnarray}}
\newcommand\eea{\end{eqnarray}}

\title{Probing bottom-associated production of a TeV scale scalar decaying to a top quark and dark matter at the LHC}

\author[a]{Amandeep Kaur Kalsi,}
\author[b]{Teruki Kamon,}
\author[c]{Seulgi Kim,}
\author[c]{Jason S.H. Lee,}
\author[b]{Denis Rathjens,}
\author[c]{Youn Jung Roh,}
\author[b(*)]{Adrian Thompson,}
\author[c]{Ian James Watson}

\affiliation[a]{Department of Mathematics, Statistics and Physics, Punjab Agricultural University, Ludhiana-141004, India}
\affiliation[b]{Mitchell Institute for Fundamental Physics and Astronomy, Department of Physics and Astronomy, Texas A\&M University, College Station, TX 77843, USA}
\affiliation[c]{Department of Physics, University of Seoul, Seoul 02504, Republic of Korea}
\affiliation[(*)]{Now at Northwestern University}

\preprint{
\begin{minipage}{5cm}
\begin{flushright}
MI-HET-803
 \end{flushright}
\end{minipage}
}

\abstract{
A minimal non-thermal dark matter model that can explain both the existence of dark matter and the baryon asymmetry in the universe is studied. It requires two color-triplet, iso-singlet scalars with $\mathcal{O}$(TeV) masses and a singlet Majorana fermion with a mass of $\mathcal{O}$(GeV). The fermion becomes stable and can play the role of the dark matter candidate. We consider the fermion to interact with a top quark via the exchange of QCD-charged scalar fields coupled dominantly to third generation fermions. The signature of a single top quark production associated with a bottom quark and large missing transverse momentum opens up the possibility to search for this type of model at the LHC in a way complementary to existing monotop searches.\newpage}

\begin{document}

\maketitle

\section{Introduction}

The puzzle of the observed matter-antimatter asymmetry in the universe has been a primary catalyst in the search for physics beyond the standard model of particle physics. 
In particular, models that violate baryon number $B$ are necessary, since this violation is among the three Sakharov conditions \cite{Sakharov:1991} necessary to accommodate the observed baryon asymmetry, the other two being $CP$-violation and the thermal inequilibrium. 
This leaves $B$-violation a robustly-motivated property of many beyond-the-Standard Model (BSM) models and searches for those at the Large Hadron Collider (LHC).

Baryon asymmetry and the existence of Dark Matter (DM) may be connected through the fact that the energy densities of DM and baryonic matter in the universe differ by only a factor of 5 -- a small difference in light of the separation of energy and mass scales that have already been observed in nature~\cite{Arkani-Hamed:2000ifx, PhysRevLett.82.896, DES:2018paw,Planck:2018vyg}. Using this coincidence of scales as a theoretical guide, concrete models that support both the observed DM relic densities and baryogenesis have been vigorously investigated in many contexts.

In particular, BSM models with $\Delta B = 2$ operators that can lie in the TeV range of QCD charged scalars have been of interest for their connection to DM~\cite{Andrea:2011ws,Agram:2013wda,Alvarez:2013jqa,Wang:2011uxa,Kumar:2013jgb, Blanke:2020bsf}, predicting resonantly-produced monotop final states in reach of LHC experiments. This is due to their simple event topology that occurs when a single top quark is produced in association with missing transverse momentum, which could be a signature of the production of DM particles. Since DM particles with masses of around 1 GeV can be probed, these search results can be also interpreted as sensitivity to mirror neutrons and mirror sector DM~\cite{Tan:2023mpj}. The monotop signature also enables distinguishing left-chiral from right-chiral  couplings~\cite{Allahverdi:2015mha}. This type of search can help to probe the aforementioned varieties of DM models and potentially shed light on neutron-anti-neutron ($n$-$\bar{n}$) oscillations, baryogenesis, non-thermal DM production, and baryon-number-violating sectors in the broader sense (see also refs.~\cite{Alonso-Alvarez:2021qfd,Racker:2014yfa}).

Several searches have already been performed at the Tevatron and the LHC with center-of-mass energies $\sqrt{s}=1.96$ TeV~\cite{Aaltonen:2012ek}, 8 TeV~\cite{Khachatryan:2014uma,Aad:2014wza}, and $13$ TeV~\cite{Sirunyan:2018gka} to probe new physics associated with baryon number violation. 
For example, scalars ($\phi$) produced via $\bar{d}_i\bar{d}_j \rightarrow \phi \rightarrow t \psi$ are excluded for masses below 3.4 TeV in events with large missing transverse momentum and a hadronically-decaying top quark in ref.~\cite{Sirunyan:2018gka}, assuming the same couplings for $\phi$ production in $d$-$s$, $s$-$b$ and $b$-$d$ quark fusions. Here $\psi$ is a DM fermion, and $i$ and $j$ ($i \not= j$) are the flavor indices of down-type quark. 
It should be noted that the limit of 3.4 TeV is not applicable if the coupling being responsible for the $d$-$s$ fusion is zero.

In this paper, we devise a search for new scalars in single top quark events with the requirement of an associated bottom quark. Thereby, we specifically probe color-triplet scalars dominantly-coupled to third generation fermions as a benchmark test for the LHC sensitivity to baryon-number-violating DM interactions. Summarily, we explore the implications for baryogenesis, $n$-$\bar{n}$ oscillations, and the DM and baryon coincidence puzzles. Experimentally, this has the desirable property of being orthogonal in selection to prior monotop searches \cite{Aad:2014wza,Sirunyan:2018gka,ATLAS:2020yzc} that veto additional $b$-tagged jets to suppress $t\bar{t}$ background for their monotop searches.

\section{The Model and its Collider Signals}
The $\Delta B = 2$ baryon-number-violating model that we consider features two charged scalars $X_{\alpha}$, $\alpha=1,2$, which are color-triplets under $SU(3)_c$, singlets under $SU(2)_L$, and have hypercharge $+4/3$. 
This scalar allows for operators such as $X_\alpha \psi u^c_i$ and $X_\alpha^* d_i^c d_j^c$, which can support baryogenesis by interference of the tree-level and loop-level decays of $X \to d_i d_j$~\cite{Allahverdi:2010im,Allahverdi:2013mza,Allahverdi:2017edd,Dutta:2014kia}. Here $i$ and $j$ ($i \not= j$) are the quark flavor indices (the color indices are omitted). In the Weyl representation, our minimal interaction Lagrangian reads
\begin{equation}
    \mathcal{L} \supset \lambda_{\alpha i} X_\alpha \psi u^c_i + \lambda_{\alpha i j}^\prime X_\alpha^* d_i^c d_j^c + \frac{m_\psi}{2}\bar{\psi}^c \psi + \text{h.c.}
\end{equation}
where $\psi$ refers to a DM candidate and ``h.c.'' refers to the Hermitian conjugate of the preceding terms in the Lagrangian. We relabel $X_{\alpha} \to X$ for the sake of convenience. The couplings of $X$ to different generations of quarks can support a wide variety of laboratory physics signatures as well. For example, the couplings to $d$-$b$ and $\psi$-$u$ can construct an effective dimension-9 operator that can give rise to $n$-$\bar{n}$ oscillations~\cite{Allahverdi:2017edd}, though we will focus on top-quark couplings in this study and only indirectly relate to $n$-$\bar{n}$ oscillations through the down-quark couplings to $d$-$b$. In addition, the $X_\alpha^* d_i^c d_j^c$ coupling can produce dijet signatures at the LHC which have been searched for by ATLAS~\cite{Aad:2014wza} and CMS~\cite{CMS:2022usq, CMS:2016gsl}, although a dedicated analysis of the dijet signature for the model considered in this study has yet to be performed.

\begin{figure}
\centering
    \begin{tikzpicture}
        \begin{feynman}
            \vertex (0);                                         
            \vertex (0a) [right=2cm of 0];                       
            \vertex (1) [above right =1.5cm of 0a] {$b$};        
            \vertex (2) [below right =1.5cm of 0a];              
            \vertex (3) [below =2cm of 0a] {$\bar{d}$};          
            \vertex (4) [right =1.5cm of 2];                     
            \vertex (5) [right =3.7cm of 3] {$\psi$};          
            \vertex (6) [right =3.3cm of 0a];                    
            \vertex (7) [above right = 1.5cm of 6] {$b$};        
            \vertex (8) [below right = 1.2cm of 6];              
            \vertex (9) [below right = 1.5cm of 8] {$\ell^{+}$};    
            \vertex (10) [above right = 1.5cm of 8] {$\nu_{\ell}$}; 
            
            \diagram* {
            (0) -- [gluon, edge label=\(g\)] (0a),
            (2) -- [fermion, edge label=\(\bar{b}\)] (0a) -- [fermion] (1),
            (2) -- [fermion] (3),
            (2) -- [scalar, edge label=\(X\)] (4),
            (4) -- [fermion] (5),
            (4) -- [fermion, edge label=\(t\)] (6) -- [fermion](7),
            (6) -- [boson, edge label=\(W^{+}\)] (8),
            (9) -- [fermion](8) -- [fermion](10),
            };
    
        \end{feynman}
    \end{tikzpicture}
    \hfill
    \begin{tikzpicture}
        \begin{feynman}
            \vertex (0);                                         
            \vertex (0a) [right=2cm of 0];                       
            \vertex (1) [above right =1.5cm of 0a] {$\bar{b}$};        
            \vertex (2) [below right =1.5cm of 0a];              
            \vertex (3) [below =2cm of 0a] {$d$};          
            \vertex (4) [right =1.5cm of 2];                     
            \vertex (5) [right =3.7cm of 3] {$\psi$};          
            \vertex (6) [right =3.3cm of 0a];                    
            \vertex (7) [above right = 1.5cm of 6] {$\bar{b}$};        
            \vertex (8) [below right = 1.2cm of 6];              
            \vertex (9) [below right = 1.5cm of 8] {$\ell^{-}$};    
            \vertex (10) [above right = 1.5cm of 8] {$\bar{\nu_{\ell}}$}; 
            
            \diagram* {
            (0) -- [gluon, edge label=\(g\)] (0a),
            (1) -- [fermion] (0a) --  [fermion, edge label=\(b\)] (2),
            (3) -- [fermion] (2),
            (2) -- [scalar, edge label=\(X^{*}\)] (4),
            (5) -- [fermion] (4),
            (7) -- [fermion] (6) -- [fermion, edge label=\(\bar{t}\)] (4),
            (6) -- [boson, edge label=\(W^{-}\)] (8),
            (10) -- [fermion](8) -- [fermion](9),
            };
    
        \end{feynman}
    \end{tikzpicture}
    \hfill
\caption{Representative Feynman diagrams for the signal process with an additional ISR $b$ quark from gluon splitting. More anti-top quarks are produced because of valence $d$ quark in proton PDFs. Here $\psi$ is a fermionic DM candidate.}
\label{fig:Feynman_diag}
\end{figure}
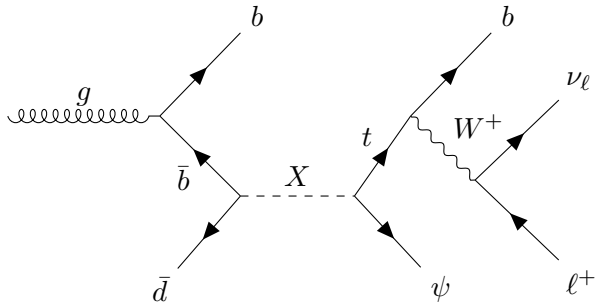

We are interested in single $t$($\bar{t}$) quark and $X$($X^{*})$ production from $\bar{d}$-$\bar{b}$($d$-$b$) and $\bar{s}$-$\bar{b}$($s$-$b$) fusion as in Figure~\ref{fig:Feynman_diag}. Due to sea quarks in the parton distribution functions (PDFs) being less likely to carry sufficient momentum, compared to valence quarks, the dominant production mode is anti-top quarks where a sea quark $b$ fuses with the valence quark $d$. The effect is the more pronounced, the heavier the $m_{X}$ is, leading to an approximate 3:1 ratio of $\bar{t}$ to $t$ for the masses probed in this publication.
To establish a sensitivity estimate for a dedicated study of the mono-top collider signature of this model, we take the limiting case in which $X_2$ is decoupled for practicality; to suppress the loop-level interference terms, we work with the assumption that $m_{X_1} \ll m_{X_2}$. While the case that $m_{X_1} \simeq m_{X_2}$ is more ideal to enhance the baryon asymmetry~\cite{Allahverdi:2017edd}, it is not necessary for successful baryogenesis. Instead, it is also possible to generate a consistent baryon asymmetry if the $X_2$ couplings are large enough, even if $m_{X_2}$ is a few factors larger than $m_{X_1}$, although the parameter space for a large enough asymmetry still diminishes quickly if $m_{X_2} \gtrsim 10 m_{X_1}$. We expect that as long as $m_{X_2} \gtrsim 2 m_{X_1}$, the production of $X_2$ with larger couplings would still be kinematically suppressed enough as to be phenomenologically similar to our case in which we take $X_2$ to be decoupled. Therefore the limit $m_{X_1} \ll m_{X_2}$ that we take in this study retains a meaningful connection to the model parameter space relevant for baryogenesis.

After focusing on the couplings to $t$, $s$, and $b$ quarks, the phenomenological part of the Lagrangian relevant for our associated jet search becomes simply
\begin{equation}
    \label{eqn:Lag1}
    \mathcal{L}_{\textrm{single top}} \supset \lambda_{\psi t} X \psi t^c + \lambda_{db}^\prime X^* d^c b^c + \lambda_{sb}^\prime X^* s^c b^c + \text{h.c.}
\end{equation}
To further indicate the production of $X$ through $\lambda_{db}^\prime$ or $\lambda_{sb}^\prime$ experimentally, we require an additional initial state radiation (ISR) $b$ quark from gluon splitting as shown in Figure~\ref{fig:Feynman_diag}. \\
Relabeling the coupling matrix indices for this specific production mode: 
$\lambda_{db}^\prime$ = $\lambda_{sb}^\prime$ $\to \lambda_{1}$,
and $\lambda_{\psi t} \to \lambda_{2}$ for the sake of clarity,  
the cross section for $X$ production with $\mathcal{O}$(TeV) masses is determined by $\lambda_{1}$, $\lambda_{2}$ and the decay width of $X$, $\Gamma_X$ as follows:
\begin{equation}
    \label{eqn:xsec1}
    \mathcal{\sigma} \propto \frac{|\lambda_{1}|^2|\lambda_{2}|^2}{\Gamma_X} 
\end{equation}
In the case of heavy mediator, the cross section is proportional to:
\begin{equation}
    \label{eqn:xsec2}
    \mathcal{\sigma} \propto \frac{|\lambda_{1}|^2|\lambda_{2}|^2}{2|\lambda_{1}|^2+|\lambda_{2}|^2}
\end{equation}
While there is a vertex factor from the gluon splitting suppressing this production mode, the steep falloff of the bottom PDF prefers a gluon parton initial state instead for heavy $X_{1}$ fixing $m_{X_2}$, lending itself to branching fraction of 24.8\% for a mass of 1 TeV, 25.2\% for 1.5 TeV, and 25.6\% for 2 TeV at $(\lambda_{1}, \lambda_{2}) = (0.1, 0.1)$.


\begin{figure}[!htbp]
\centering
\includegraphics[width=0.53\linewidth]{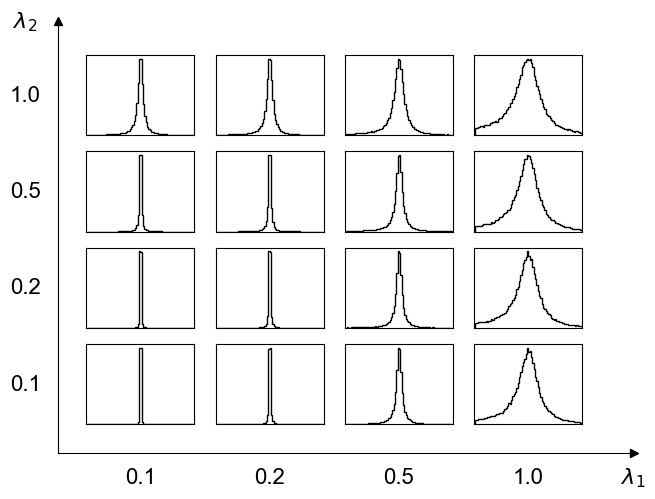}
\caption{Mass distribution at truth-level for $X_1$ as a function of ($\lambda_{1}, \lambda_{2}$) for $m_{X_1} = 1$ TeV. The decay width is proportional to $2|\lambda_{1}|^2+|\lambda_{2}|^2$. The $x$-axis of each subplot is in the range of [800, 1200] GeV. All the subplots are normalized to the total event in each case.}
\label{fig:MX1_xsec_monotop}
\end{figure}

\begin{table}[]
    \centering
    \begin{tabular}{c|c|c|c|c}
    \hline\hline
    \multicolumn{5}{c}{cross section $\sigma$ [fb]} \\
    \hline
    \multicolumn{1}{c|}{\multirow{2}{*}{$\lambda_1$}} &     \multicolumn{4}{c}{$\lambda_2$} \\
    \cline{2-5}
    \multicolumn{1}{l|}{} & 0.1 & 0.2 & 0.5 & 1.0 \\
    \hline
    \multicolumn{1}{c|}{0.1} & 2.22 & 5.42 & 9.12 & 10.3 \\
    \multicolumn{1}{c|}{0.2} & 2.40 & 8.13 & 25.5  & 36.3 \\
    \multicolumn{1}{c|}{0.5} & 2.47 & 9.62 & 51.1  & 132 \\
    \multicolumn{1}{c|}{1.0} & 2.57 & 10.3 & 61.1 & 210 \\
    \hline\hline
    \end{tabular}
  \caption{Cross sections for $m_{X_1} = 1$ TeV and $m_{X_2} = 10$ TeV.  The cross section is following Equation \ref{eqn:xsec2}.}    
    \label{tab:xsec}
\end{table}

\section{Simulation}
\label{sec:simulation}

\texttt{MadGraph5\_aMC@NLO (v2.6.7)} \cite{Alwall:2014hca} has been used for the generation of the single top quark plus bottom quark signal events, as well as all background samples listed from this point onward, in proton-proton collisions at $\sqrt{s}=13$\,TeV. As we include diagrams of different total jet multiplicities with $b$ quarks directly from the proton PDF or from gluon splitting, we utilize 5-flavor PDF definition and MLM matching.
Parton showering and hadronization were performed by \texttt{PYTHIA8 (v8.244)}  \cite{SJOSTRAND2015159}.

Up to two additional ISR/final state radiation (FSR) jets, excluding one ISR $b$ jet and one FSR $b$ jet, are included in the signal generation. The analysis is restricted to leptonic top-quark decays ($t \to Wb \to l \nu b$), where $l$ is a muon or an electron.
The mass of $X_1$ is set to 1 TeV, 1.5 TeV or 2 TeV, of $X_2$ to 10 TeV, and of $\psi$ to 1 GeV. Samples were generated with an $11\times 9$ grid of $\{\lambda_{1}, \lambda_{2}\}$ with $\lambda_{1}$ in range of [0.01, 1.0] and $\lambda_{2}$ in range of [0.05, 1.0] for $m_{X_{1}}$ = 1 TeV, and an $11\times 5$ grid of $\{\lambda_{1}, \lambda_{2}\}$ with $\lambda_{1}$ in range of [0.01, 1.0] and $\lambda_{2}$ in range of [0.1, 1.0] for 1.5 and 2 TeV. We set the upper limit to 1.0 to constrain the width of $X$ within the resonant regime. 
The distributions of $m_{X1}$ as a function of $\{\lambda_{1}, \lambda_{2}\}$ are shown in Figure~\ref{fig:MX1_xsec_monotop}.



We have generated background samples for $W(\to l \nu)$+jets (up to two additional jets), $Z(\to l^{+} l^{-})$+jets (up to two additional $b$ jets), single top quark, semileptonic $t \bar{t}$, and diboson ($WW$, $WZ(\to l \nu b\overline{b})$, $ZZ$) events. Hadronic $t\bar{t}$ is neglected because the size of the contribution is a few percent of semileptonic $t\bar{t}$ after applying selections described in the next sections.
We generated one million events for each single top quark, $WW$ and $ZZ$ sample. Other backgrounds were generated in 8 bins of $H_\textrm{T}$ with one million events per bin. All samples were simulated at leading order (LO). 

A CMS-like detector simulation was performed using \texttt{DELPHES (v3.4.2)}\cite{deFavereau:2013fsa} with a modified \texttt{DELPHES} card. In the leptonic top quark decays, the lepton is difficult to resolve due to the boost of the top quark from the $X$ decay causing the lepton to be included in the bottom jet during the jet clustering. Therefore, we have modified the standard CMS \texttt{DELPHES} card to remove isolation. The anti-$k_{\textrm{T}}$ jet algorithm \cite{Cacciari_2008} is used with jet radius $R$ = 0.4 to identify jets. The $b$-tagging efficiency has been also updated by emulating the efficiency for the medium working point of the CSVv2 algorithm \cite{CMS-PAS-BTV-15-001,BTV-16-002}. 




\section{Object and event selection}
\label{sec:selection}

We define in this chapter a set of baseline selections and object definitions before applying any machine-learning, when the section is explicitly not about machine learning.

In signal events, we expect to have one $b$-tagged jet originating from gluon splitting on the one hand, and a top quark decaying to another $b$-tagged jet, one lepton ($e$ or $\mu$) and a neutrino on the other hand. Additionally, a large amount of missing transeverse momentum $p_{\textrm{T}}^{\textrm{miss}}$ is expected, due to the DM candidate $\phi$ and the neutrino.
Therefore, we select events with at least two $b$-tagged jets. They are required to have ${|\eta| <}$ 2.4 and $p_{\textrm{T}} >$ 50(30) GeV for the (sub-)leading $b$-tagged jet. Additional jets are allowed. Furthermore, we select $p_{\textrm{T}}^{\textrm{miss}} >$ 50 GeV and one light lepton ($e$ or $\mu$) with $p_{\textrm{T}} >$ 30 GeV and ${|\eta| <}$ 2.5(2.4) for electrons (muons) without isolation requirements.
We expect the top quark from the $X$ decay to be boosted in our model, as $X$ is far heavier than the top quark. This often causes the leading lepton to be clustered into the leading $b$-tagged jet. As the proportion of transverse momentum within typical jets carried by light leptons is small (typically less than $5\%$, see e.g. \cite{CMS:2016lmd}), we don't expect distinguishing the high $p_{\textrm{T}}$ leptons from most non-prompt leptons to be an issue. To avoid misjudging the top-quark-originated leading $b$-tagged jet's direction and overestimating its momentum, we subtract the highest $p_{\textrm{T}}$ lepton in the event from its jets, whenever it is a constituent of the jet. We define leading $b$-tagged jet $b_{1}$ and sub-leading $b$-tagged jet $b_{2}$ after this subtraction, where their $p_{\textrm{T}}$ values satisfy our baseline selection of 50 and 30 GeV, respectively.

\begin{figure}[!htbp]
\centering
\includegraphics[height=0.33\linewidth]{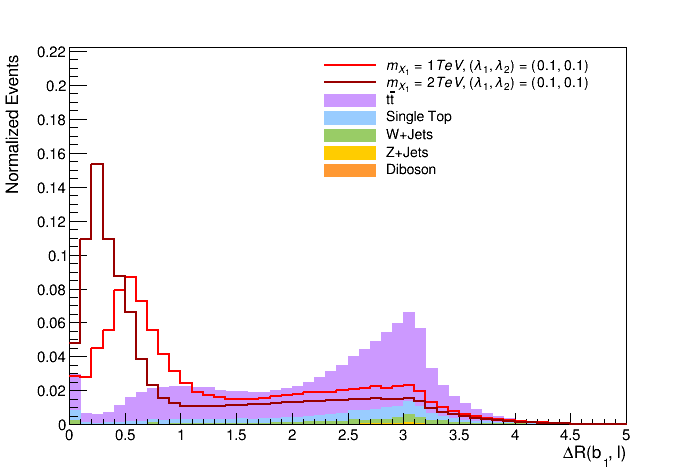}
\includegraphics[height=0.33\linewidth]{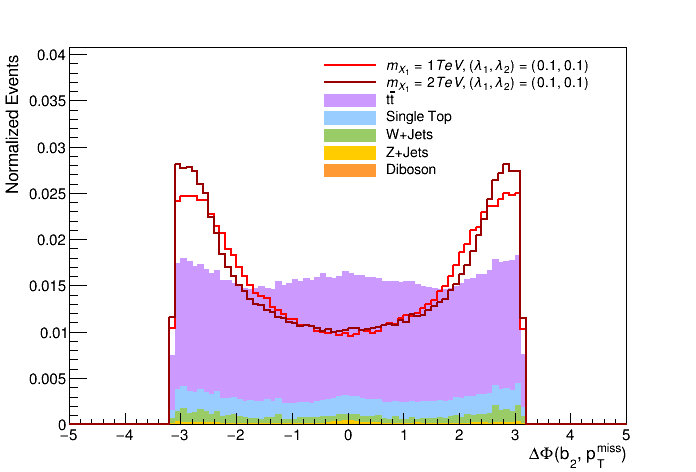}
\caption{$\Delta R$($b_1$, $l$) (left) and $\Delta \phi(b_2, p_{\textrm{T}}^{\textrm{miss}})$ (right) distributions after applying baseline selections. We expect that the leading $b$-tagged jet ($b_1$) and the selected lepton are from the top decay and $b_2$ is from gluon splitting. As the top quark is more boosted, $\Delta R$($b_1$, $l$) becomes smaller. Moreover, $b_2$ likely appears in back-to-back to $p_{\textrm{T}}^{\textrm{miss}}$ where it represents a vector sum of momenta of the neutrino from the top quark decay and the DM particle. The lowest $\Delta R$($b_1$, $l$) bin of background is mostly from secondary leptons carrying a large fraction of the jet momentum.
}
\label{fig:MT_monotop}
\end{figure}

The $\Delta R(b_1, l)$ distribution in Figure~\ref{fig:MT_monotop} shows that the leading lepton points at the rim of the jet when it is clustered unlike high-momentum leptons from hadronization. They tend to originate in the center of jet, where most of the energy during hadronization is deposited. Thus, the resulting angular difference from distinguishing the boosted lepton has good background separation power. Not only $b_{2}$ and $X$ are back-to-back due to momentum conservation, but $X$ is essentially at the rest in the laboratory frame, so its decay products also have maximal separation in $\Delta \phi$. Consequently, as $p_{\textrm{T}}^{\textrm{miss}}$ is dominated by the $\phi$ and neutrino contributions, it also exhibits a large angular distance in the transverse plane to $b_{2}$, as shown in Figure~\ref{fig:MT_monotop} and serves well for background separation. These effects are the more pronounced, the heavier $X_1$ is.

The significance with baseline selections is 0.11 (0.016, 0.0033) for $m_{X_{1}}$ = 1 (1.5, 2) TeV sample at $(\lambda_{1}, \lambda_{2}) = (0.1, 0.1)$. In this analysis, significance is defined as ${n_s / \sqrt{n_b}}$, where $n_s$ and $n_b$ are the number of signal and background events, respectively.

\section{BDT training}

Boosted decision trees (BDTs), provided by the Toolkit for Multivariate Data Analysis with \texttt{ROOT} \cite{Hoecker_2007}, are used to combine the discriminating power of several kinematic distributions into a single classifier in order to further improve the separation between signal and backgrounds.
The BDT training is performed for events after our baseline selections in signal samples with $\lambda_{1} = \{0.1, 0.3, 1.0\}$ and $\lambda_{2} = \{0.1, 0.3, 1.0\}$ and the background samples mentioned in Section~\ref{sec:simulation}.
A separate BDT was trained for each signal mass point with the combined signal samples.

\begin{figure}[!htbp]
\centering
\includegraphics[height=0.37\linewidth]{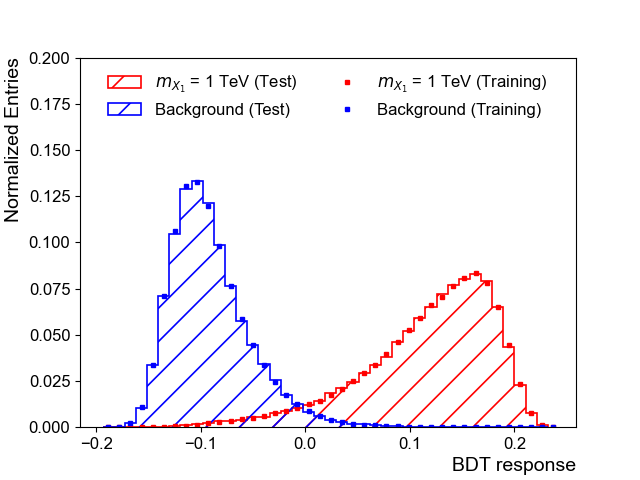}
\includegraphics[height=0.37\linewidth]{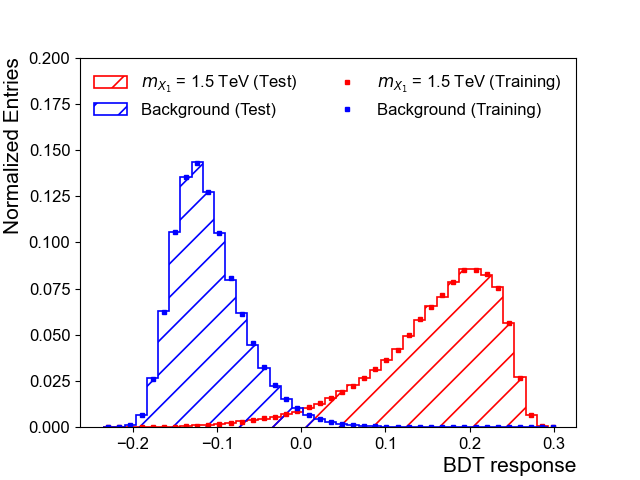}
\includegraphics[height=0.37\linewidth]{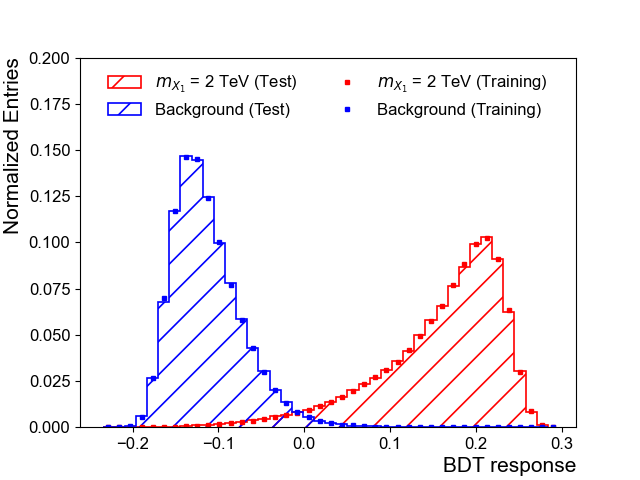}
\includegraphics[height=0.37\linewidth]{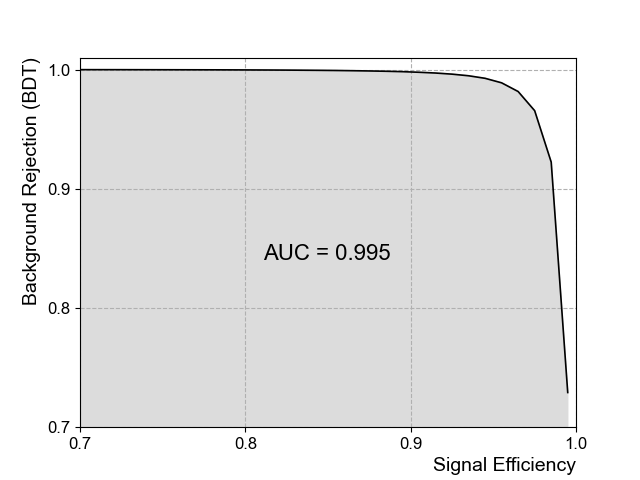}
\caption{The model response of the BDT for $m_{X_{1}}$ of 1 (top left), 1.5 (top right), and 2 TeV (bottom left) case and the ROC curve of the BDT for $m_{X_{1}}$ of 2 TeV (bottom right) with the optimized BDT hyperparameters.}
\label{fig:BDT_monotop}
\end{figure}

The following input variables were used in training: $p_{\textrm{T}}$ and $\eta$ of the $b_1$, $b_2$, $l$, and $b_1+l$, $\Delta R(b_1, b_2)$, $\Delta R(b_1, l)$, $\Delta \phi(b_2, p_{\textrm{T}}^{\textrm{miss}})$, $\Delta \phi(l, p_{\textrm{T}}^{\textrm{miss}})$, $m_\textrm{T}(b_1, p_{\textrm{T}}^{\textrm{miss}})$, and $m_\textrm{T}(l, p_{\textrm{T}}^{\textrm{miss}})$. 
To get the best performance, an optimization of the BDT hyperparameter on \texttt{NTree} and \texttt{MaxDepth} was conducted. (NTrees, MaxDepth) = (3000, 5) for $m_{X_{1}}$ = 1 TeV and (2500, 5) for 1.5 TeV and 2 TeV were chosen as the optimized parameters. 
The variable with the largest variable importance value was $\Delta \phi(b_2, p_{\textrm{T}}^{\textrm{miss}})$, as shown in the right plot of Figure~\ref{fig:MT_monotop}. The model response and receiver operating characteristic (ROC) curve of the BDT with the optimized hyperparameters are shown in Figure~\ref{fig:BDT_monotop}, which shows that the separation improves as $m_{X_1}$ increases. The area under the curve (AUC) of the ROC curve for $m_{X_1}$ = 1 (1.5, 2) TeV is 0.989 (0.994, 0.995).

\begin{figure}[!htbp]
\centering
\includegraphics[height=0.37\linewidth]{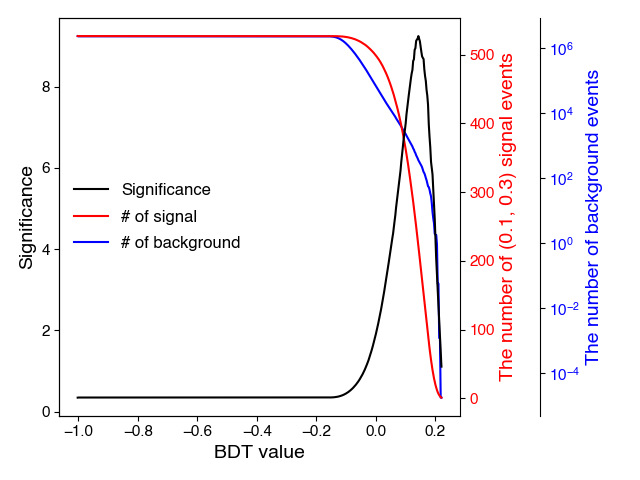}
\caption{Significance and the number of events remaining as a function of the BDT discriminator with $m_{X_{1}}$  = 1 TeV, and $(\lambda_{1}, \lambda_{2}) = (0.1, 0.3)$.}
\label{fig:Significance_monotop}
\end{figure}

In addition to the baseline selections as defined in Section~\ref{sec:selection}, we require that events have BDT $> \alpha$, where $\alpha$ is optimized to produce the largest signal significance.
We chose the value of $\alpha$ as 0.1425 (0.1900, 0.2100) for $m_{X_1}$ = 1 (1.5, 2) TeV scenario. Figure~\ref{fig:Significance_monotop} shows significance and the number of events remaining versus $\alpha$ for $(\lambda_{1}, \lambda_{2}) = (0.1, 0.3)$, $m_{X_1}$ = 1 TeV signal samples.

\section{Results}
We evaluate the expected exclusion limits on three benchmark scenarios with $m_{X_{1}}$ = 1, 1.5, and 2 TeV in our model on $\{\lambda_{1}, \lambda_{2}\}$ space for each mass point, assuming integrated luminosity scenarios of 300 fb$^{-1}$ and 3000 fb$^{-1}$ corresponding to the projected Run 3 and Phase 2 LHC luminosities, respectively. The feasibility is computed in the $\{\lambda_{1}, \lambda_{2}\}$ space.
\texttt{RooStats} \cite{Moneta:2010pm} is used to perform the statistical analysis. We create a model based on Poisson statistics 
\begin{equation}
    \label{eqn:Pos}
        f(N_{b}; N_{s}+N_{b}) = \frac{{(N_{s}+N_{b})^{N_{b}}}e^{-(N_{s}+N_{b})}}{N_{b}!}
\end{equation}
while considering the statistical uncertainty as Gaussian-distributed, taking
\begin{equation}
    \label{eqn:Gaus}
        g(N_{b}) = e^{-\frac{(N_{b}-N_{b_{0}})^2}{2{\sigma_b}^2}}
\end{equation}
Here, $N_{s}$ is the expected number of signal events, $N_{b}$ is the expected number of background events which has an uncertainty of $\sigma_{b}$ and $N_{b_0}$ is the estimated number of background events after applying the baseline selections and cut on BDT output variable. The model probability density function is defined as:
\begin{equation}
    \label{eqn:PROD}
        f(N_{b}; N_{s}+N_{b})\cdot g(N_{b})
\end{equation}
We used a uniform distribution as a prior probability density function on the number of signal events. Our Monte Carlo (MC) statistical uncertainties are 10\%-30\%, which is of a similar magnitude as CMS results~\cite{CMS:2018gbj}. In the main interpretation, we did therefore not consider additional systematic uncertainties. We explored instead the worsening of limits when adding a flat 10\% systematic normalization uncertainty. At 300 fb$^{-1}$, this added uncertainty would worsen the expected 95\% CL limit for $\lambda_{2}$ at $\lambda_{1}>$ 0.15 for $m_{X_{1}}$= 1 (1.5, 2) TeV by 33 (14, 3.2)\%, indicating that higher masses would be expected to be statistically-dominated, while systematic improvements would primarily affect lower mass scenarios.

\begin{figure}[!htbp]
\centering\includegraphics[width=0.47\linewidth]{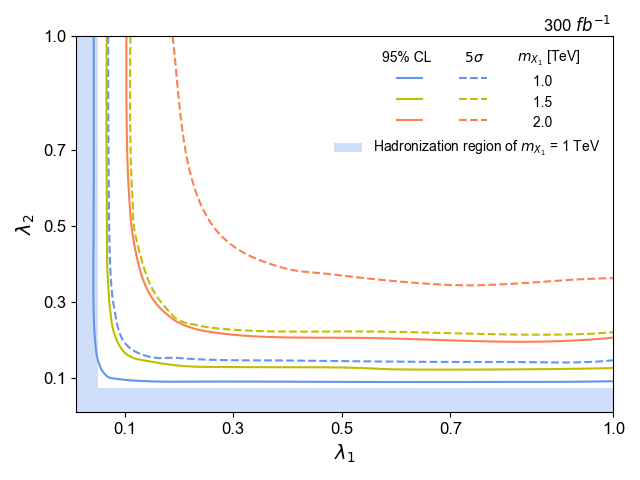}
\includegraphics[width=0.47\linewidth]{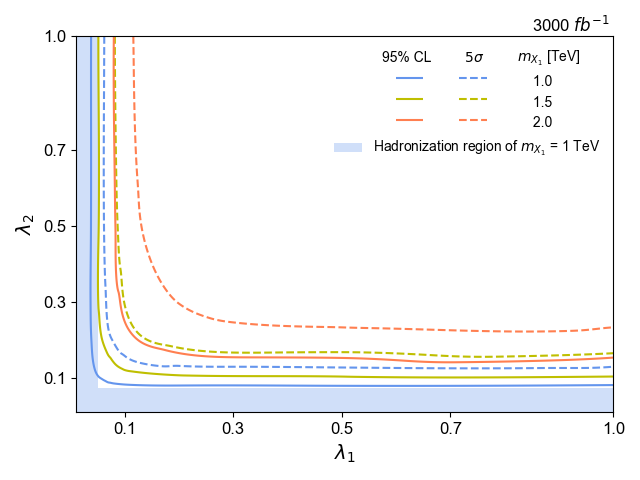}
\caption{Limits on $\{\lambda_{1}, \lambda_{2}\}$ at 95\% confidence level and 5$\sigma$ discovery potential for $m_{X_{1}}$ = 1.0, 1.5 and 2.0 TeV at 300 fb$^{-1}$ (left) and 3000 fb$^{-1}$ (right). Regions above the contours are expected to be excluded.}
\label{fig:limit_monotop}
\end{figure}

Limits on $\{\lambda_{1}, \lambda_{2}\}$ are computed using \texttt{RooStats} BayesianCalculator. It computes the posterior probability based on the Poisson likelihood in Equation~\ref{eqn:Pos}. Figure~\ref{fig:limit_monotop} shows the 95\% confidence level exclusion and the 5$\sigma$ discovery potential of $X_1$ with masses of 1, 1.5, and 2 TeV for integrated luminosities of 300 fb$^{-1}$ expected by the end of Run 3 and 3000 fb$^{-1}$ expected by the end of High Luminosity LHC era. 
At 300 fb$^{-1}$, we expect this type of analysis could feasibly constrain the coupling related to the top quark, $\lambda_{2}$, to be less than 0.09 (0.12, 0.21) for most of the parameter space in the $m_{X_{1}}$ = 1 (1.5, 2) TeV cases at 95\% CL. For the coupling related to the down-type quarks, $\lambda_{1}$, at 300 fb$^{-1}$ a similar constraint can be made to be less than 0.04 (0.07, 0.11). At 3000 fb$^{-1}$, $\lambda_{2}$ is constrained to be less than 0.08 (0.11, 0.15), and $\lambda_{1}$ is constrained to be less than 0.04 (0.05, 0.08). The limits have a rectangular form with rounded edges, so these simple relations do not hold for when both $\lambda_{1}$ and $\lambda_{2}$ approach their respective stated borders, simultaneously.



At the hadronization scale, the non-perturbative effects are expected to be large, leading to the unreliable calculation from \texttt{MadGraph5\_aMC@NLO}. This lower limit of $\lambda_i$ following perturbative QCD theory bounds its lifetime within the hadronization scale. 
We marked the region of $m_{X_{1}}$ = 1 TeV which suffers from this issue by a solid area in Figure \ref{fig:limit_monotop}. The lower limits of $\lambda_i$ of the other two scenarios are not visualized in Figure~\ref{fig:limit_monotop} since it doesn't affect the limit. 


Higher masses of $X_1$ are more challenging to exclude, as the cross section decreases, albeit the branching fraction of additional $b$-quarks from ISR gluon splitting has a trend in the opposite direction as tree level contributions of the $b$ PDF diminish faster with increasing energy requirements than the gluon PDF scaling does, while the additional vertex factor for the splitting remains constant.
The limits presented here will be improved in combination with an analysis of events with hadronically-decaying top quarks, probing the interesting region of $\lambda_i \approx {\cal O}(0.1)$ \cite{Allahverdi:2017edd}. In the same reference, a bound on $|\lambda_1|$ by interpretation of an ATLAS dijet cross section measurement~\cite{ATLAS:2017ble} and a light dijet search~\cite{CMS:2016gsl} is made to the effect of $|\lambda_1| \lesssim 0.1$ at $m_X = 1$ TeV, indirectly. A dedicated study of dijet final states that include b-tagged jets \cite{PhysRevD.98.032016}~would offer a more direct constraint, or a study to interpret constraints from more recent 13 TeV data~\cite{CMS:2022usq}, which has yet to be performed.

\section{Summary}

We have investigated the potential of searching for baryon-number-violating DM candidates in association with a bottom jet, missing transverse momentum, and a top quark at the LHC. The DM candidate in this model is a singlet Majorana fermion that interacts with top quarks via the exchange of QCD-charged scalar $X_1$. The applicability of the analysis is limisted to light DM particles ($m_\phi << m_{X_1}$).
The additional ISR $b$ quark was required to distinguish production modes of $X_1$ via $\lambda_{db}^\prime$ or $\lambda_{sb}^\prime$
experimentally from previous search strategies.
The 95\% upper limits and $5\sigma$ limits on couplings are shown for $m_{X_1} =$ 1, 1.5, and 2 TeV at an integrated luminosity of 300 fb$^{-1}$ expected by the end of Run 3 and 3000 fb$^{-1}$ by the end of High Luminosity LHC era.
This reach is based on a novel approach requiring an associated $b$-tagged jet and utilizing machine-learning techniques. Traditional monotop analyses veto additional $b$-tagged jets to suppress backgrounds. We therefore show that the inclusion of the second $b$-tagged jet offers a complementary analysis design with comparable reach to search for DM candidates with baryon-number-violating interactions. These are of particular importance, given they could explain both the observed DM abundance as well as the baryon asymmetry in the universe, simultaneously. Our analysis utilized a leptonic top quark decay while e.g. the 13 TeV CMS analysis~\cite{Sirunyan:2018gka} utilizes boosted hadronic top quark identification, so there is further room given the larger branching ratio of the hadronic decay for possible improvement with another dedicated hadronic analysis.

\section*{Acknowledgements}

We thank B. Dutta for the motivating discussions and guidance in the model selection and search strategy. 
T.K. D.R. and A.T. are supported in part by the US Department of Energy grant DE-SC0010813. S.K. and J.L. are supported by the National Research Foundation of Korea (NRF) grant No. 2023R1A2C2002751.
Y.R. and I.W. are supported by NRF grants  No.2021R1A2C1093704 and 2018R1A6A1A06024977, respectively.

\bibliographystyle{JHEP}
\bibliography{main}

\end{document}